# An SME's Adoption of a Cloud Based Integrated Management System (IMS) When Certifying against Management System Standards (MSS)

**Ming Hock Yew**
School of Business
SIM University
Singapore
Email: mhyew001@unisim.edu.sg

**Jenson Chong-Leng Goh**
School of Business
SIM University
Singapore
Email: jensongohcl@unisim.edu.sg

## Abstract

This case study introduces a four step approach used by a Singapore small and medium enterprise (SME) in implementing a cloud computing based integrated management system (IMS) to meet ISO 9001, ISO 14001, and OHSAS 18001 certification requirements. The objectives of this case study are to study: (1) the challenges encountered by an SME during the IMS integration process at each of the four levels of integration (2) the extent which Frugal IT Innovation and Technology Acceptance Model (TAM3) concepts apply to our four step approach. The four step approach was assessed against a framework of four integration levels of the management systems (MS), and also the applicability of two IS theories (Frugal IT Innovation and TAM3). Data was collected via: (1) direct observations (2) participant observations; (3) interviews with key personnel involved in the project; and (4) analysis of documents pertaining to the project. This case study provides an exemplary model on using IS theories and public cloud technologies to develop an effective IMS that other SMEs can learn from.

## 1 Introduction

A management system standard (MSS) is defined as a set of standardized requirements for a management system (MS) which details how an organization is to be designed and managed (Furusten 2000). An MSS provides a framework for best practices, processes, and procedures in a specific business activity that an organization can adopt in order to fulfil its business objectives, leading to a significant improvement in its process efficiency and a reduction in its information-related transaction costs (Nadvi et al. 2004). In a globalized business competitive environment, the endorsement by a certification body that an organization has met all the requirements of an MSS can further '*simulate international trade by eliminating obstacles arising from different national practices*' (Heras-Saizarbitoria et al. 2013, pp. 48-49). As a result of these benefits, there has been an explosive growth of MS adoptions in organizations across the world since the early 90s, with statistics (ISO 2011) suggesting an increasing trend, even amongst the small and medium enterprises (SMEs), to pursue multiple certifications. Today, three of the most popular MSSs are ISO9001 (quality management system - QMS), ISO14001 (environmental management system - EMS), and OHSAS (occupational health and safety management system - OHSMS). They are established by standards organizations such as the International Standards Organization (ISO).

Due to the need to avoid overlaps in processes, procedures and documentation in multiple MSs, the industries have been progressively moving towards implementing integrated management system (IMS) in order to leverage on the synergies derived from such implementation (Casadesus et al. 2011). For instance, it was reported that 85% of organizations surveyed had sought aggressively to integrate their MSs (Karapetrovic et al. 2006). While Gianni et al. (2015) allured to a lack of international, world-wide recognized standard for developing an IMS , there had in fact recently been efforts along these lines by the industries and the MSS bodies. The ISO body has just revamped new versions of the MSSs (e.g. ISO 9001 : 2015 and ISO 14001 : 2015) to provide "*identical structure, text and common terms and definitions for MSS of the future*" (International Standards Organization, ISO 2015) . This "*high level structure*" is a set of guideline for all future ISO MSSs and will ensure consistency in implementation among future and revised MSSs. Thus, making the integrated use of these MSSs a simpler process.

As the research and business community has long recognized the potential of cloud computing in levelling the playing field for SMEs (Lacity et al. 2014), we see an opportunity for SMEs to leverage on cloud computing in implementing an IMS. Even with extant research on SME adoption of cloud computing, research on IMS implementation via cloud computing by SMEs is non-existent. We therefore adopt two suitable models to assess the adoption of cloud computing in implementing an IMS by SMEs, namely the frugal IT innovation by Ahuja et al. (2014) and the Technology Adoption Model (TAM) by Davis (1980). Leveraging upon the extant literature in IMS, cloud computing and IT adoption, this research presents, using an illustrative case study of a Singapore SME, a practice oriented four step methodology to implement a successful public cloud computing based IMS to fulfil the ISO 9001, ISO 14001, and OHSAS 18001 MS requirements.

A case study approach is adopted for this research as the SME's adoption of cloud computing for its IMS implementation is an uncommon instance of such phenomenon amongst SMEs. Furthermore, a case study approach allows an in-depth study of the SME's IMS development and implementation, and as such could potentially lead to generalization in such a practice for other SMEs in a similar situation. In so doing, we believe we have contributed both to the cloud computing and IMS literature and also provide a viable roadmap that can guide SMEs on how to leverage on public cloud technologies to successfully implement an IMS. The ensuing section presents our literature search on the challenges and benefits of implementing IMS via cloud computing by SMEs. This is followed by our methodology for this research, our results from the study, and discussions on the results, lessons learnt and recommendations, as well as conclusions to the study.

## 2 Literature Review

### 2.1 Benefits and challenges of MS certification by SMEs

The development of an IMS is particularly advantageous to an SME because it not only helps the SME reduces the associated overheads to maintain multiple sets of documents and processes required by different MSSs, but also assists the SME to improve its process efficiency in various management areas such as quality, environmental, safety, and business continuity concurrently. Despite these associated benefits of implementing an IMS, it is common for SMEs to choose not to adopt any IMS as they are often constrained by resources. Even for those who choose to adopt an IMS, they are often faced with several challenges in assimilating the new IMS into its day to day operations. These key challenges faced by an SME during the implementation include: (1) the lack of internalization of IMS's processes by the organization's employees; (2) the high associated cost of implementing IMS; and (3) the excessive amount of documentation required by each MSS within an IMS (Boiral 2011).

Furthermore, research has shown that if the motivation for implementing an IMS is due to requirements imposed externally on an organization, there is a good chance that the implementation of it would be viewed as a barrier to innovation by employees, which predictably will lead to its failure in its implementation (Castillo-Rojas et al. 2012). However, if the emphasis of implementing an IMS is on improving business performance, such implementation will be viewed favourably by employees and would likely result in its successful implementation (Castillo-Rojas et al. 2012). Hence, it is imperative for an organization to develop an effective organization change strategy to convince its employees when attempting to adopt an IMS.

While these challenges have been well-documented in the extant literature (e.g. Boiral 2011; Simon et al. 2012), there is little research that attempts to provide a processual model through the use of cloud computing to resolve these identified challenges. This is a missed opportunity as the associated benefits of implementing an effective IMS are significant to any organization especially in acquiring market recognition after being certified to the ISO9001 and ISO14001 (Simon et al. 2012).

### 2.2 Overcoming IMS challenges by focusing on its integration levels

The main hurdle in the implementation of an effective IMS comes from the challenges during the process of integration across the organization (Karapetrovic 2003; Karapetrovic et al. 1998). For instance, based on a comprehensive review of the extant literature by Simon et al. (2011), the authors identified four key challenges during the process of developing an IMS and they are: (1) lack of resources for integration; (2) difficulties with the MS implementation and certification; (3) organizational internal difficulties; and (4) difficulties with the people working with the MSSs. These challenges are particularly salient and are difficult to overcome by SME. From a survey of 76 organizations that implemented an IMS, Simon et al. (2012)) highlighted four integration levels that SME can target to overcome all/some of the identified challenges above: (1) Strategic and operating procedures; (2) Control procedures; (3) Documentation resources; and (4) Human resources (HR).

## 2.3  SME IS Adoption and Adoption Theories

Given the pervasive need for information management and the high reliance on documentation for successful IMS implementation, it would be logical to assume that IS would add significant value in the adoption of an IMS in an organization. While there has been much research on the usefulness of IS in many disciplines, there is little research on the use of IS to address the challenges encountered when implementing an IMS. We posit that the use of IS in the implementation of an IMS is essential as it can induce this emphasis on improving business performance in the mind of the employees, which is critical to its success (Castillo-Rojas et al. 2012). This is because we think that many processes and procedures as advocated by the mainstream MSSs can be automated and streamlined through the appropriate use of IS. With such automation, employees will perceive the implementation of an IMS as an attempt to improve business performance and this will lead to higher acceptance by employees of the change and a better business performance especially in an SME (Castillo-Rojas et al. 2012).

In the adoption of IS, SMEs have long been challenged by stretched IS management capabilities, few technically skilled employees, tight access to financial capital and few slack resources. Because of these challenges, SMEs often have slower adoption rates for IT innovations than larger firms (Levy et al. 2000). As such, SMEs generally lag behind on the use of IS to facilitate the process of implementing an effective IMS as compared to large organizations. Both IS researchers and practitioners, however, have long taken the stand that cloud computing services "*have the potential to alleviate the historical information systems (IS) challenges facing SMEs*" (Lacity et al. 2014, pp31)  The emergence of cloud-based technologies and services appeared to have levelled the playing field for SMEs, particularly in adopting new business practices such as IMS. While the adoption and diffusion of cloud computing as an IS innovation have been extensively studied, there is still much interest in cloud computing for SMEs, given their low and patchy adoption rate (Alshamaila et al. 2013). This is particularly so with respect to the use of cloud computing services for IMS by SMEs.

A suitable framework to assess the adoption of cloud computing for IMS by an SME as an innovation is the frugal IT innovation, conceptualized as a combination of business, social and technology innovation by Ahuja et al. (2014). While the resource based view (RBV) theory was considered for such studies, it is deemed to have its limitations where competitive markets and adoption of IS innovations are concerned. RBV assumes that "*firms can be conceptualized as bundles of resources. By doing so, it misidentifies the locus of long-term competitive advantage in dynamic markets, overemphasizes the strategic logic of leverage, and reaches a boundary condition in high-velocity markets*" (Eisenhardt et al. 2000, pp1105). RBV hence has been criticised as being "*tautological and lacking relevance in dynamic business environments*", with frugal IT innovation seen as more realistic for firms with limited resources, such as *SMEs* (Ahuja et al. 2014, pp8). Four principles outlined in frugal IT innovation (Ahuja et al. 2014) are: (1) Seek opportunity in adversity; (2) Do more with less; (3) Think and act flexibly; (4) Include the margin. While the first three principles are self-explanatory, principle 4 relates to engaging local communities and partners to locally build, deliver, and support their solutions — making these solutions in turn affordable, accessible, and sustainable (Ahuja et al. 2014).

Adoption of technological innovations in organizations (large or small) led to concerns of resistance or outright rejection in some cases. To address this, technology adoption (or acceptance) research had yielded established models (TAM, TAM2, TAM3) which were validated and laid the groundwork for further research for adoption of a specific technology. Davis (1989b) developed TAM based on two core constructs of perceived ease of use (PEOU) and perceived usefulness (PU) which represents users perception of the technology. Adoption of a particular technology is predicted by these two user's perceptions. Subjective norm (SN) was added into TAM, leading to TAM2 (Venkatesh et al. 2003). In TAM3 PU and PEOU have four antecedents namely, individual differences, system characteristics, social influences and facilitating conditions. This allows organization to craft strategies that will induce acceptance of the technology that has just been rolled out. Given its applicability, we leverage on TAM3 in our assessment of acceptance of the public cloud computing based IMS for the SME in our case study.

## 2.4  Objectives

Guided by our literature review the two objectives of this case study are to study: (1) the challenges encountered by an SME during the IMS integration process at each of the four levels of integration as highlighted by Simon et al. (2012); (2) the extent which frugal IT innovation and TAM3 concepts apply to our four step approach of implementing a cloud computing based IMS.

# 3   Methodology

With the counsel of an MS consultant and inputs from cloud computing service providers, the development of the cloud computing based IMS was devised over these four distinct steps and they are listed as follows. Data collection was by way of: (1) direct observations (2) participant observations, where one of the authors participated in IMS meetings; (3) interviews with key personnel involved in the project (MD, production manager, production staff, sales staff, and cloud computing vendor) ; and (4) analysis of documents pertaining to the project. Data analysis was carried out by systematically searching the data to identify and/or categorize specific observable actions or characteristics.

### Step 1: Review individual MSS requirements to identify areas of integration in these requirements

Firstly, MSS requirements are determined via the clauses in the MSS and grouped according to the four integration levels (strategic and operating procedures; control procedures; documentation resources; and HR).

### Step 2: Translate integrated requirements to IS requirements

A key technique used here was to determine the flow of information based on the integrated requirements identified in step 1. From the nature of the information required, the requirements for the IS are determined. The eight steps of IMS development process proposed by Gianni et al. (2015) (i.e. process mapping, documentation, training, internal audits, management review, preventive and corrective actions, external audits, and certification) are adopted here for the IMS development.

### Step 3: Evaluate and select cloud computing solutions to match IS requirements

In this step, the suitability of various cloud computing options for the type of IS identified in step 2 is evaluated based on the types of cloud computing services and technologies offered. Before selecting a particular cloud computing service or technology, companies need to first decide on the type of cloud computing service that is required. The National Institute for Standards and Technology (NIST), USA categorizes cloud computing services into four types - public, private, community, and hybrid. The most wide-spread type of cloud computing service, public clouds, can be further categorized into Infrastructure as a Service (IaaS), Platform as a Service (PaaS), and Software as a Service (SaaS). To aid evaluation, criteria for selecting the type of cloud computing service includes cost, availability, security requirements, scale, and range of end-users.

### Step 4: Deploy the cloud computing based IS solution

Deployment involves three main steps: (1) Configuring the cloud computing service; (2) Migrating existing on premise infrastructure, applications and data to the new cloud computing service; and (3) Managing the change process and training of end-users.

# 4   An SME's Cloud Computing IMS Implementation Journey

## 4.1   SME Profile and Initial Management Decision on Cloud Computing

We present a short profile of our SME before the detailed analysis of each of the four steps of the cloud computing based IMS implementation. Anson Interiors Private Limited (Anson)[1] is a Singapore based office system furniture design and manufacturing SME. It has a showroom and a factory in Singapore as well as another factory in neighbouring Malaysia, and about a hundred employees. A significant proportion of Anson's customers are from the governmental sector. Selling to these customers requires Anson to demonstrate that it is certified to ISO 9001, ISO 14001 and OHSAS 18001 concurrently. Given the limited resources available to Anson, the senior management decided that the development of a public cloud computing based IMS is the most cost-effective approach. This information was obtained from participant observations when one of the authors sat in a steering committee (comprising of top management) meeting. The decision to use public cloud computing was due to its perceived most cost effectiveness. This underlines the application of principle 1 in frugal IT innovation

---

[1] The company name and names of interviewees are masked as requested by the senior management of the company

- seek opportunity in adversity. Further case study data collection and analysis are presented via the four step approach adopted by Anson in its IMS development and implementation.

## 4.2 Step 1: Review individual MSS requirements to identify areas of integration in these requirements

Developing an IMS begins with an understanding of the requirements as stated in the clauses of each of the component MSS that makes up the IMS. We illustrate this step with an example based on ISO 9001 and ISO 14001 MS development. To guide the identification of the common requirements of the MSSs, clauses from both MSSs are grouped based on the four levels of integration (Simon et al. 2012) as tabulated in table 1 below. Grouping the requirements of both MSSs in such a manner helped Anson's management plan the IMS by determining the processes and resources required.

| **ISO9001** | **ISO14001** |
|---|---|
| **Strategic procedures** | |
| 4.1. General requirements | 4.1. General requirements |
| 5.1. Management commitment | 4.2. Environmental policy |
| 5.2. Customer focus | 4.3. Planning |
| 5.3. Quality policy | |
| 5.4. Planning | |
| 7.1. Planning for product realization | |
| 7.3.1. Design and development planning | |
| **Operating procedures** | |
| 6.1. Provision of resources | 4.4.7. Emergency preparedness and response |
| 6.3. Infrastructure | |
| 6.4. Work environment | |
| 7.2.1. Determination of requirements related to the product | |
| 7.2.3. Customer communication | |
| 7.3.2. Design and development inputs | |
| 7.3.3. Design and development outputs | |
| 7.4.2. Purchasing information | |
| 7.5.4. Customer property | |
| 7.5.5. Preservation of product | |
| 8.4. Analysis of data | |
| 8.5.1. Continual Improvement | |
| **Control procedures** | |
| 5.5.3. Internal communication | 4.4.3. Communication |
| 5.6. Management review | 4.4.6. Operational control |
| 7.2.2. Review of requirements related to the product | 4.5 Checking (except 4.5.4) |
| 7.3.4. Design and development review | 4.6. Management review |
| 7.3.5. Design and development verification | |
| 7.3.6. Design and development validation | |
| 7.3.7. Control of design and development changes | |
| 7.4.1. Purchasing process | |
| 7.4.3. Verification of purchased product | |
| 7.5.1. Control of production and service provision | |
| 7.5.2. Validation of processes for production and service provision | |
| 7.5.3. Identification and traceability | |
| 7.6. Control of monitoring and measuring equipment | |
| 8.2 Monitoring and measurement | |
| 8.3. Control of nonconforming product | |
| 8.5.2. Corrective action | |
| 8.5.3. Preventive action | |
| **Documentation resources** | |
| 4.2. Documentation requirements | 4.4.4. Documentation |
| | 4.4.5. Control of documents |
| **Human resources** | |
| 5.5. 1. Responsibility and authority | 4.4.1. Resources, role, responsibility and authority |
| 5.5.2. Management representative | |
| 6.2. Human resources | 4.4.2. Competence, awareness and training |

| | 4.5.4. Control of records |
| --- | --- |

*Table 1. Summary of ISO9001 and ISO14001 clauses grouped by integration level requirements*

As the MSs' requirements tend to overlap, they provide avenues for integration at the procedural levels (strategic, operating and control). To verify integration at these three levels, Anson's IMS documentation was reviewed. An integrated manual on strategic procedures ensure that requirements at the strategic level by the MSSs that make up the IMS are met. Like-wise a common set of documents pertaining to operating and control procedures was used to guide operations and control in order to meet the requirements of all the standards within the IMS. From an interview with a production staff, one feedback obtained that emphasizes the importance of integration of documentation resources was *"ISO documents seem foreign to us and we are worried about the additional documentation requirements"*.

As an SME with limited human resources, it was observed that Anson did not maintain dedicated human resources for each of the MSS. Instead, each personnel in charge of a certain aspect of the IMS is responsible for all the MSSs with respect to that aspect. In both ISO 9001 and ISO 14001 MSSs, the roles for top management are to: (1) determine the high level requirements of the MSs (as stipulated by the MSSs) and plan for its development; (2) communicate (internally and externally) the MSs requirements and their importance; (3) establish policies (a quality policy and an environmental policy in this case), roles, responsibilities and authorities with respect to the MSs; (4) ensure availability of resources for the MSs; and (5) conduct management reviews of the MSs. This suggests integration of HR at the highest level, and was demonstrated when Anson's managing director (MD) performed all of these five tasks for the three MSSs within the IMS. The high level management involvement (aided by the systematic approach to determine common MSS requirements) addresses difficulties with the MS implementation and certification, as well as organizational internal difficulties. This demonstrates adoption of principle 3 of frugal IT innovation - Think and act flexibly. At the operational level, the requirements for HR were also found to be common for both MSS, as outlined in clauses 5.5 and 6.2 of ISO9001 and 4.4 of ISO14001, indicating further avenues for integration at the HR level. Here again, a review of Anson's organization chart reveals such integration with a conspicuous example of HR department fulfilling MS requirements for competency development across the MSSs.

.

## 4.3   Step 2: Translate integrated requirements to IS requirements

By reviewing all the clauses of both standards, it was found that in a QMS (ISO 9001), information flows between customers (external party) and the organization when the customer articulates its requirements and when the measurement of customer's satisfaction takes place (e.g. Voice of Customer). Like-wise in an EMS (ISO14001), the communications with external parties on environmental management is with stakeholders such as regulatory bodies, customers, vendors and environmental special interest groups (SIGs). There are, therefore, strategic, operating, control, and documentation procedures concerning information management and requiring integration here. Internally, the information flows between and within the MS processes.

Both standards require the organization to exert control of its processes and maintain updated documented procedures and records, as stipulated in clauses 4 to 7 (ISO 9001) and clauses 4.1 to 4.6 (ISO 14001). There were therefore efforts to "*Include the margin*" (principle 4 of frugal IT innovation) practised here. With this view of the information flow, the types of IS that can be used to support the processes are determined, as tabulated in table 2 below. Leveraging on the IS theory (Chen et al. 2008) that articulates the role of IS (automate, informate and transform), the integration levels are also mapped to the type of IS and its role as shown.

| Integration level | Type of IS required | IS roles |
| --- | --- | --- |
| Strategic and operating procedures | Document management system to manage manuals (which contains the documented MS, plans and policies) and operating procedures | *Automate* document management process |
| | Messaging systems (including emails) to facillitate communication (internally and externally) | *Automate* the communication process and *informate* stakeholders |
| | Enterprise systems (or document management | *Automate* work processes and |
| | system for control of records if enterprise system not available | *informate* where data (or records) are generated |
| | Collaboration systems (e.g. video conferencing, online bulletins, social media, blogs | *Automate* and *transform* collaboration processes |
| | Functional systems for functional processes | *Automate* and *transform* functional processes |
| Control procedures | Document management systems to manage documented procedures and records | *Automate* document and record management process |
| | Enterprise systems where data is generated as records for control purposes | *Automate* the control process and *informate* users |
| Documenation resources | Document management systems to manage documented procedures and records | *Automate* document and record management process |
| Human resources | Document management system to manage competence, training and awareness documents and records. | *Automate* and transform HR management |

*Table 2. Integration level requirements of ISO9001 and ISO14001 and the types of IS that can support it*

Based on an interview with the cloud computing vendor, five types of ISs that can support the integration level of the IMS as identified in Table 1 are: (1) document management system; (2) messaging system; (3) collaboration system; (4) enterprise system, and (5) functional systems. In this phase (based on data from interviews with the MD) the steering committee also uncovered several complex integration requirements of an IMS that would require the support of a specialized IS. From a minute of the IMS project meeting, this can be seen when each of the requirements of both MSSs was reviewed against Anson's business processes, as tabulated in Table 3. The functional ISs identified from this review are: (1) Learning Management System (LMS); (2) Computer Aided Design (CAD) system; (3) Asset information management system; (4) Data analytics system; and (5) Workflow management system. From the range of ISs, it can be seen that a document management system could provide IS support for the majority of the IMS processes in Table 3. Such a system was used to manage both the documents and records mandated by ISO9001 and ISO14001. At the same time, a messaging system (in particular email), while necessary, is already ubiquitous. Collaboration systems that can go beyond the commonly available video conferencing were also found to be of value. Such systems could also incorporate document editing collaboration, e.g. co-authoring during video conferencing.

| ISO 9001 : 2008 Clauses | | ISO 14001 : 2004 Clauses | | Type of IS to Support MS Processes that are complex |
|---|---|---|---|---|
| No | Clause | No | Clause | |
| 6.2.2 | Competence, training and awareness | 4.4.2 | Competence, training and awareness | Learning Management Systems (LMS) if huge and complex HR |
| 7.3 | Design and development | | | Computer Aided Design (CAD) systems. Necessary when design intensive work is encountered. |
| 7.6 | Control of monitoring and measuring instrument | 7.6 | Monitoring and measurement | Asset information management systems (AIMS) if assets are of high value and complex |
| 8.2.3 | Monitoring and measurement of process | 4.5.1 | Monitoring and measurement | Process control systems if very process intensive. |
| 8.2.4 | Monitoring and measurement of product | | | |
| 8.4 | Analysis of data | | | Data analytics systems if huge amount of data is encountered and analytical requirement are high. |
| 8.3 | Control of non-conforming product | 4.5.3 | Nonconformity, corrective action and preventive action | Workflow management systems if workflow is complex |
| 8.5.2 | Corrective action | | | |
| 8.5.3 | Preventive action | | | |

*Table 3. Complex requirements of ISO9001 and ISO14001 that needs to be supported by specialized IS*

With an agenda of adopting the most cost efficient IS for the job (principle 2 of frugal IT innovation - Do more with less) , the steering committee felt that a reliable document management system would be able to meet most of Anson's IS requirements to implement an effective IMS. The steering committee also felt that a CAD system that allows collaboration amongst its designers and sales team, will increase its employee's productivity. As administering nonconformities, corrective actions and preventive actions are workflow intensive activities; the steering committee felt that a workflow management system is also necessary to meet the IMS requirements identified. In sum, this analysis has narrowed the ISs required for Anson's IMS down to the following systems: (1) Document management system; (2) Messaging system; (3) Collaboration system; (4) CAD system; and (5) Workflow management system. This allows the steering committee to move on to next phase of evaluating the type of IT infrastructure that best serves Anson's IS requirements.

### 4.4 Step 3: Evaluate and select cloud computing solutions to match IS requirements

The types of IS fleshed out in the preceding section is summarized below:

- Document management system
- Messaging system
- Enterprise system
- Collaboration system
- LMS
- CAD
- AIMS
- Process control systems
- Data analytics management system
- Workflow management system

Of these, Anson's management had determined that the document management system, collaboration system and CAD system are key systems. This is based on the integration level they could achieve. Of the four types of cloud computing service (public, private, community, and hybrid), public cloud was deemed to be the best option. This is based on the criteria of cost and availability criteria. As the type of IS required are all software applications, the type of pubic cloud selected was SaaS.  To be sure of the cost advantage, Anson's steering committee compared the cost of deploying on premise IT infrastructure and through public cloud computing services. The criteria for comparisons are based on the key architectural requirements and the total cost of ownership (TCO) of each proposal. Request for quotations (RFQ) were sent out to vendors that supply on premise IT infrastructure solutions and the accompanying application software solutions and to vendors that supply cloud computing services. The comparison shows that on premise IT infrastructure requires a significant amount of architectural components, in stark comparison to the corresponding SaaS which only requires the application software on the cloud to be owned (or leased) by the SaaS user. At the same time, both upfront cost and annual recurring cost of the integrated SaaS solution was significantly lower than that of an on premise solution. This swung the steering committee's decision towards the SaaS based solution, further demonstrating adherence to frugal IT innovation principle 2 (Do more with less).  The task ahead was then narrowed down to selecting the most suitable public cloud solution for Anson.

Two integrated SaaS solutions were found to offer a suite of solutions that included document management, messaging, collaboration and workflow management. These were Google App and Microsoft Office 365 (O365). Both were suites comprising of the following application software that matches the IS requirements as defined in Phase 2 and they are summarized as follows: (1) Application software (spreadsheet, word processing and presentation) known collectively as desktop apps; (2) Cloud based email, calendaring and contacts management; (3) Document management system; (4) Configurable workflow management system. The steering committee evaluated the two cloud technologies based on the following criteria.

*First Criteria: Compatibility with existing desktop apps used*. Like the majority of businesses, Anson uses Microsoft Office suite for spreadsheet, word processing, and presentation. On the one hand, O365 works natively with Microsoft Office documents and allows such documents to be created and edited over the browser. This increases perceived ease of use among employees if O365 is chosen (Davis 1989a). On the other hand, Google Docs (the desktop app in Google App) appears to lack all the editing features when it is being used to edit Microsoft Office documents. There was also feedback that the Microsoft Office file edited through Google Docs might lose some document features. This raised concerns that the Google Docs file format was not readily acceptable by Anson's customers and staff that use Microsoft Office at work in general, hence O365 met this criteria better.

*Second criteria: Document management capabilities and security*. Google Drive (part of Google App) does not provide a secured centrally controlled document repository as compared to the Sharepoint Online (part of O365). Instead, users share their storage space with other users. Key requirements for an IMS is its ability to support document and record control (clauses 4.2. for ISO 9001 and 4.4.4., 4.4.5. for ISO 14001). This requires the technology to support access control, approval routing, and document version control - features that are available on O365, but not completely on Google Docs.

*Third criteria: Good integration between desktop app, messaging, calendaring, contacts management, video conferencing and document management to facilitate collaboration among employees*. This is an important criterion for Anson given that its employees are located across three different work sites. Both O365 and Google App provides for these integration.

*Fourth criteria: Configurable workflow management features*. Sharepoint Online (O365) allows configurable workflow management, hence supporting clauses such as corrective and preventive action (8.5.2, 8.5.3 in ISO 9001 and 4.5.3 ISO 14001), which is another plus point for O365.

While the evaluation suggests that O365 is the preferred solution, the steering committee wanted to ensure that the company's employees felt likewise. Hence they further surveyed key employees and determined that O365 was the appropriate choice.

At the same time, with no other options to cater to the specialized IS requirements, Autocad 365 (for CAD) was selected. As the employees have been using AutoCAD (desktop) for some times, the steering committee do not feel that there will be an issue in transiting to this new platform. While there was a need for a cloud computing based enterprise system (CRM and ERP) to support the IMS, the steering committee felt that this, being a large project, could be deferred till a later stage after the IMS is successfully implemented.

## 4.5   Step 4: Deploy the Solution

The cloud computing based solution deployment was carried out with the help of service providers specializing in the O365 and Autocad 365 deployment. During the deployment process, the Technology Acceptance Model 3 (TAM3) was referred to extensively (Venkatesh et al. 2008) to ensure the successful adoption of the deployed public cloud technologies by employees. With the advice of the consultant, the steering committee contemplated the four antecedents of TAM3 to determine how these influences can be used as change levers to increase employee's acceptance. *"We are sure that implementing an IMS based on cloud technologies is the way to go, except that we are not too sure what is the best way to get our people to be happy with these changes and use these cloud based solutions effectively."* --- Managing Director, Anson Interiors Pte Ltd.

To create the right social influences that encourage adoption under the TAM3, maximizing user participation during the implementation process was considered. The company achieved the participations of users through (1) instilling strong leadership and accountability throughout the deployment process; (2) providing constant communication; and (3) delegating specific tasks related to system implementation to be performed directly by the users. These efforts had helped to induce the importance of the change towards public cloud and rally the employees behind these changes. These efforts allowed employees to become more familiar with the system characteristics. To create a good facilitating condition that will moderate PU and PEOU, the company aggressively provided comprehensive organizational support during the change process. This involves the implementation of a comprehensive set of training for employees, the development of strong end-user support capabilities, and the publication of reference material on how to use the public cloud to support the IMS. To address the challenges of individual differences and system characteristics, cloud awareness communication and training on the use of O365 and Autocad 365 for the IMS was carried out. The main theme for communication was "*Cloud is the new norm*". O365's document management and collaboration tools were used for the training and staff was allowed to access the training material (videos, presentation and guides) on demand. Online guides were also posted up with summarized steps to access these pasted on every workstation. All of these change management processes further demonstrated application of principle 3 of frugal IT innovation (Think and act flexibly).

## 5　Discussions

The four challenges during the process of integration in an IMS development were mainly addressed in the cloud computing based IMS implementation of Anson Interiors Pte Ltd. The systematic approach in step 1 of reviewing MSS requirements based on the four levels of integration by Simon et al. (2012) to determine common requirements led to high levels of integration for the procedures (strategic, operating and control), documentation resources and HR. This addressed the challenges of standards implementation and certification, organizational internal difficulties, and difficulties with the people working with the standards. Identification of the type of IS required to support the IMS (step 2) as well as the subsequent evaluation/selection (step 3) and deployment of a cloud based IMS (step 4) automates a significant portion of the IMS work. This reduces reliance on labour and addresses the challenge of lack of resources for integration. The entire project took fourteen months to complete and this was a little longer than the original timeline of one year, as there were initial apprehension on migrating to the cloud based IMS. The steering committee decided to allow for a longer implementation period so as to ensure full acceptance strategies of the system. All four principles of frugal IT innovation were applied in the development and implementation of the IMS.

Management reviewed the effectiveness of the SaaS based IMS at the end of the implementation based on the following criteria: (1) Quality improvement; (2) Environmental performance improvement; (3) Cost avoidance over on premise solutions; and (4) Productivity improvement. Quality was measured based on customer satisfaction level and the informal survey. A sales person commented that his customer is pleased to have quicker response to their queries, thanks to the better coordinated customer servicing as a result of the use of O365 and AutoCAD 365 collaboration tools. Cost avoidance as compared to on premise solutions was estimated to be SGD$110,000 and an annual cost saving of SGD$19,000. Operations cost savings due to productivity improvement was determined from the reduction in average man-hours spent processing each dollar value of business transaction. This was found to have dropped from 0.0020 man-hours per dollar to 0.0011 per dollar over a six month period after the implementation of O365, i.e. an almost 100% increase in productivity where handling transactions is concerned. This translates to an annual direct cost savings of SGD$45,000 from the use of O365. The challenges of organizational internal difficulties and difficulties with the people working with the standards were hence addressed with the automation of the key IMS processes via cloud computing, (i.e. document management system) and with strong leadership.

## 6　Lessons Learnt & Recommendations

As part of the post implementation review, the steering committee summarized the following key lesson learnt as follows:

Firstly, it was agreed that the IMS integration requirements needs to be tailored to the SME's business processes and then systematically translated into IS requirements before selecting the appropriate cloud computing solution. Most SME may jump right to cloud computing solutions without going through this process. The fit between business requirements, IS requirements and technology is extremely important for a complex project like the implementation of an IMS. Here, the roles of IS fleshed out in table 2 demonstrates how this fit was achieved. A good example was the document management system that not only facilitated document and record control, but also supported most of the other MS processes. An Anson's staff commented on how he was greatly encouraged when they realized that the traditional way of printing and circulating controlled copies of the documented procedures and manuals was not required with the new cloud computing based IMS. *"I never realized we could do without any print-outs of the controlled copies of the documented procedures --- Anson's production staff.* Anson's success in its IMS implementation could be reviewed against the success factors for IMS implementation by Boiral (2011). The company had shown great managerial conviction (1st success factor) by ensuring comprehensive communication to its employees and the company had also listened to the issues raised by its staff. Staff understood the reasons for certification (2nd success factor) via the consultation and communication sessions. A large part of the steering committee's work was in mobilizing the employees progressively (3rd success factor). This management led implementation of the IMS had ensured that the requirements of both MSS were internalized within the company (4th success factor). The advantages of a cloud based IMS had convinced the staff to adopt this technology and hence allowed integration of the IMS with the goals of the MSSs (4th success factor).

From these lessons learnt, the recommendations for SMEs aspiring to implement a cloud based IMS would be to: (1) consider adopting the four step approach in tandem with the four levels of integration as outlined in our study; (2) ensure a comprehensive understanding of the MSSs and the affected organization's business processes before the consideration of IS requirements; (3) involve the middle management and executives in the early stages of the project and consider a proof of concept to be performed by employees; (4) leverage on the TAM3 model to create the necessary environment that is necessary to facilitate the adoption of the technologies across the organization; and (6) apply the four principles of frugal IT innovation.

## 7  Conclusion & Limitations

The approach for successfully implementing a cloud computing based IMS based on ISO9001 , ISO14001 and OHSAS 18001 MSSs had been shown to be viable in our study. We have detailed a comprehensive discussion on how our case organization address the challenges faced in such implementation and provided insights on how to address them through the combinative use of public cloud technologies and IS theories. We think our study is of significant value to practitioners implementing multiple management systems, possibly beyond the two staple MSS, i.e. ISO 9001 and ISO 14001. As the study involves only one SME, it will not be possible to generalize the findings and solutions applied in this SME across all SMEs. This represents a limitation to the study and also opens up possibilities for future research work. Another limitation is the non-exhaustive review of SaaS solutions for the integrated MS. The quality and environmental performance improvement, cost savings and productivity gains enjoyed by the SME, however, provide strong indication that a cloud computing based SaaS could effectively replace on premise IT infrastructure and fulfil the role of an IS to support an IMS.

## Copyright